\documentstyle[12pt,titlepage]{article}

\font\grande=cmr9.5 scaled \magstep4
\font\medio=cmr9.5 scaled \magstep2
\outer\def\beginsection#1\par{\medbreak\bigskip
      \message{#1}\leftline{\bf#1}\nobreak\medskip
\vskip-\parskip
      \noindent}

\def\laq{\raise 0.4ex\hbox{$<$}\kern -0.8em\lower 0.62
ex\hbox{$\sim$}}
\def\gaq{\raise 0.4ex\hbox{$>$}\kern -0.7em\lower 0.62
ex\hbox{$\sim$}}

\begin{document}
\bibliographystyle {unsrt}

\titlepage

\begin{flushright}
CERN-PH-TH/2004-132
\end{flushright}

\vspace{15mm}
\begin{center}
{\grande Homogeneous and isotropic big rips? }\\
\vspace{15mm}
 Massimo Giovannini \footnote{Electronic address: massimo.giovannini@cern.ch}\\

\vspace{0.3cm}
{{\sl Centro ``Enrico Fermi", Compendio del Viminale, Via 
Panisperna 89/A, 00184 Rome, Italy}}\\
\vspace{0.3cm}
{{\sl Department of Physics, Theory Division, CERN, 1211 Geneva 23, Switzerland}}
\vspace*{1cm}

\end{center}

\vskip 2cm
\centerline{\medio  Abstract}

\noindent
We investigate the way big rips are approached  in a fully 
inhomogeneous description of the space-time geometry. 
If the pressure and energy densities are connected by a 
(supernegative) barotropic index,  the spatial gradients and the 
anisotropic expansion decay as the big rip is approached. This 
behaviour is contrasted with the usual big-bang singularities.
A similar analysis is  performed in the case 
 of sudden (quiescent) singularities and it is argued that 
 the spatial gradients may well be non-negligible in the vicinity of 
 pressure singularities. 
\vspace{5mm}

\vfill
\newpage

Consider the situation where the  matter content of the present Universe 
is dominated  by a perfect fluid with barotropic index $ w = p/\rho < -1$.
Evidence of this possibility seems to be suggested from the 
analysis of Type Ia supernovae. If this is the case future 
singularities may be expected \cite{sta1,rip1} (see also Refs. \cite{rip2,rip3,rip4}).  

In the present paper we intend to study the nature of future big rip 
singularities in a fully inhomogeneous approach whose 
relevance in the context of usual big-bang singularities 
has been exploited long ago \cite{BK1} (see also \cite{f2,mgio2} and references therein). 
For instance, in the case of conventional big-bang singularities one 
can show that the relative contribution of the gradients decays 
as the singularity is approached. Does the same happens in the 
case of future (big-rip) singularities? In the case of big-bang singularity 
the anisotropy is believed to play an important r\^ole in the 
way the singularity is effectively approached. is this true also 
for big-rips? These are some of the questions we ought to address.

Consider first the case where the perfect barotropic fluid 
filling the Universe is characterized by a 
supernegative equation of state, i.e. 
$ w = - 1 - \epsilon$, with $\epsilon > 0$.  We shall then be interested 
in the contribution of the spatial gradients as the the big rip 
is approached. To achieve this goal Einstein equations 
must be written in fully inhomogeneous terms. 
By writing the line element as \footnote{Note that, in this approach, 
$\gamma_{ij}(\vec{x},t)$ contains $6$ independent degrees of freedom 
corresponding to the correct number of initial conditions required for a 
general discussion of the problem.}
\begin{equation}
ds^2 = dt^2 - \gamma_{ij}(\vec{x}, t) dx^{i} dx^{j},
\end{equation}
the Hamiltonian and momentum constraints take the form \footnote{In the following the overdot will denote 
a partial derivation with respect to the cosmic time coordinate.}
\begin{eqnarray}
&& K^2 - {\rm Tr}K^2 + r = 16\pi G [(p + \rho) u_{0} u^{0} - p],
\label{HAM}\\
&& \nabla_{i} K - \nabla_{k} K^{k}_{i} = 8\pi G ( p + \rho) u^{0} u_{i},
\label{MOM}
\end{eqnarray}
where 
\begin{equation}
K_{i}^{j} = - \frac{1}{2}\gamma^{j k} \frac{\partial }{\partial t} \gamma_{k i}, 
\qquad K = K_{i}^{i},\qquad {\rm Tr} K^2 = K_{i}^{j} K_{j}^{i},
\label{EC}
\end{equation}
and where $r = r_{i}^{i}$ is the trace of the spatial (instrinsic) curvature 
computed from the three dimensional Ricci tensor in terms of $\gamma_{ij}$.
The $(ij)$ components of Einstein equations are, instead, 
\begin{equation}
\frac{1}{\sqrt{\gamma}} \frac{\partial}{\partial t} \biggl(\sqrt{\gamma} K_{i}^{j} \biggr) - r_{i}^{j} = 4\pi G [ - 2 (p + \rho) u_{i} u^{j} + (p - \rho) \delta_{i}^{j} ].
\label{ij}
\end{equation}
where $\gamma = {\rm det}(\gamma_{ij})$. 

Consider then the following expansion of the spatial metric, i.e. 
\begin{equation}
\gamma_{ij}(\vec{x}, t) = a^2(t) [ \alpha_{ij}(\vec{x}) + \beta_{ij}(\vec{x},t)],
\label{MT}
\end{equation}
when the term $\beta_{ij}(\vec{x},t)$ contains the contribution of the gradients 
while $\alpha_{ij}(\vec{x})$ does not contain any gradient. Recalling that the 
inverse metric, to this order in the gradient expansion, is given by 
$\gamma^{ij} = [ \alpha^{ij} - \beta^{ij}]/a^2(t)$, the extrinsic 
curvature (and its traces) can be readily computed. For instance, from Eqs. 
(\ref{EC}) and (\ref{MT})
\begin{equation}
K_{i}^{j} = - (H \delta_{i}^{j} + \frac{\dot{\beta}_{i}^{j}}{2}), 
\label{EC2}
\end{equation}
where $H = \dot{a}/a$.
Since from the momentum constraint (\ref{MOM}) 
the velocity field is always of higher order in the 
gradient expansion, i.e. 
\begin{equation}
u^{0} u_{i} = \frac{\partial_{k} \dot{\beta}_{i}^{k} - \partial_{i}
\dot{\beta}}{16\pi G ( p + \rho)},
\end{equation}
the contribution of the peculiar velocity field can be 
neglected in the remaining equations. This is not true 
necessarily to higher order in the gradient expansion.

Thus, recalling that $u_{0}u^{0} = 1 + \alpha^{ij} u_{i} u_{j}/a^2(t)$, and 
using Eqs. (\ref{EC})--(\ref{EC2}),
Eqs. (\ref{HAM}) and (\ref{ij}) can be written, respectively, as 
\begin{eqnarray}
&& 6 H^2 + 2 H \dot{\beta} + \frac{{\cal P}}{a^2} = 16\pi G \rho,
\label{HAM1}\\
&& 2(\dot{H} + 3 H^2) \delta_{i}^{j} + \ddot{\beta}_{i}^{j} + 3 H \dot{\beta}_{i}^{j} 
+ H \dot{\beta}\delta_{i}^{j} + \frac{2}{a^2} {\cal P}_{i}^{j} =  8\pi G ( \rho - p) \delta_{i}^{j},
\label{ij1}
\end{eqnarray}
where we defined $r_{i}^{j} = {\cal P}_{i}^{j}/a^2$ (in this 
notation ${\cal P}_{i}^{j}$ and its traces are time-independent).

Clearly the trace-free part of Eq. (\ref{ij1}) reduces to 
\begin{equation}
\biggl(\ddot{\beta}_{i}^{j} - \frac{1}{3} \ddot{\beta} \delta_{i}^{j} \biggr) 
+ 3 H \biggl( \dot{\beta}_{i}^{j} - \frac{1}{3} \dot{\beta} \delta_{i}^{j}\biggr) = 
- \frac{2}{a^2}\biggl({\cal P}_{i}^{j} - \frac{1}{3} \delta_{i}^{j} {\cal P} \biggr). 
\label{TF}
\end{equation}
 Equation (\ref{ij1}) allows to determine the gradient 
 contribution to the energy density and the following relation 
 \begin{equation}
 3( 2 \dot{H} + 3 H^2) + \ddot{\beta} + 3 H \dot{\beta} + \frac{{\cal P}}{2 a^2} = 
 -24\pi G p,
 \label{pressure}
 \end{equation}
 allows to determine the gradient contribution to the pressure 
 density. 
 
 Even if not strictly necessary we can imagine 
 to split $\rho$ and $p$ as 
 \begin{equation}
 \rho = \overline{\rho} + \tilde{\rho},\qquad p = \overline{p} + \tilde{p},
 \label{rhop}
 \end{equation}
 where, from Eqs. (\ref{ij1}) $\overline{\rho}$ and $\overline{p}$ obey the usual 
 Friedmann equations 
 \begin{eqnarray}
 && H^2 = \frac{8\pi G}{3} \overline{\rho},
 \label{F1}\\
 && ( 2 \dot{H} + 3 H^2) = - 8\pi G \overline{p},
 \label{F2}\\
 && \dot{\overline{\rho}} + 3 H (\overline{\rho} +\overline{p}) =0.
 \label{F3}
 \end{eqnarray}
Eq. (\ref{F3}) comes from the $(0)$ component 
of the covariant conservation equation, i.e. 
\begin{equation}
\frac{1}{\sqrt{\gamma}} \frac{\partial}{\partial t} \{ \sqrt{\gamma} [ ( p + \rho )u_{0} u^{0} - p]\} - \frac{1}{\sqrt{\gamma}} \partial_{i} [ \sqrt{\gamma} 
( p+ \rho) u_{0} u^{i} ] - K_{k}^{\ell} [ (p + \rho) u^{k} u_{\ell} + p \delta^{k}_{\ell}=0.
\label{cons1}
\end{equation}
Equation (\ref{cons1}) also implies that  $\tilde{\rho}$ and $\tilde{p}$ 
obey
\begin{equation}
\frac{\partial \tilde{\rho}}{\partial t} + 3 H ( \tilde{\rho} + \tilde{p}) + 
\frac{\dot{\beta}}{2}( \overline{\rho} + \overline{p}) =0.
\end{equation}
According to the same logic, Eq. (\ref{HAM1}) defines $\tilde{\rho}$ 
in terms of $\dot{\beta}$ and ${\cal P}$, i.e. 
\begin{equation}
2 H\dot{\beta} + \frac{{\cal P}}{a^2} = 16\pi G \tilde{\rho}.
\end{equation}
The fully inhomogeneous solution of the system can then be derived and it is: 
\begin{eqnarray}
&&\beta_{i}^{j}(\vec{x},t) = a^{-2 - 3\epsilon} {\cal B}_{i}^{j}(\vec{x}) 
\label{beta1}\\
&& K_{i}^{j} = - H \biggl[\delta_{i}^{j} - \frac{3\epsilon + 2 }{2} a^{-2 - 3 \epsilon} {\cal B}_{i}^{j}(\vec{x})\biggr].
\label{K1}
\end{eqnarray}
The space-dependent tensor  ${\cal B}_{i}^{j}$
 is related to the three-dimensional curvature tensor ${\cal P}_{i}^{j}$ by virtue
 of Eq. (\ref{ij1}); the explicit result is 
\begin{equation}
{\cal P}_{i}^{j} = - \frac{H_{0}^2}{4} [ ( 3 \epsilon + 2 ) ( 3 \epsilon -2) 
{\cal B}_{i}^{j} - ( 3 \epsilon^2 + 12 \epsilon + 4)  {\cal B} \delta_{i}^{j} ].
\label{curv1}
\end{equation}
The physical significance of Eq. (\ref{curv1}) is most easily understood 
by inverting Eq. (\ref{curv1}), i.e. 
\begin{equation}
{\cal B}_{i}^{j} = \frac{4}{H_{0}^2 ( 4 - 9\epsilon^2)}\biggl[ {\cal P}_{i}^{j} -
\frac{3 \epsilon^2 + 12 \epsilon + 4}{4 ( 9 \epsilon + 4)} {\cal P} \delta_{i}^{j}\biggr].
\label{curv2}
\end{equation}
Equation (\ref{curv2}) determines ${\cal B}_{i}^{j}$ as a function 
of the three-dimensional spatial curvature computed from the $\alpha_{ij}(\vec{x})$. The form of $\alpha_{ij}$ is in a sense 
arbitrary. But once $\alpha_{ij}(\vec{x})$ is fixed, ${\cal B}_{i}^{j}(\vec{x})$ 
follow immediately from Eq. (\ref{curv2}). 

To derive  Eqs. (\ref{curv1}) and (\ref{curv2}) the following parametrization 
for the scale factor and for the Hubble factor
\begin{equation}
a(t) = \biggl( \frac{t_{\rm br} - t}{t_{0}}\biggr)^{- \frac{2}{3 \epsilon}},\qquad H(t) = H_{0} a^{\frac{3}{2} \epsilon},
\label{HP1}
\end{equation}
has been used.  In Eq. (\ref{HP1}) $t_{\rm br}$ denotes the value 
of the cosmic time at the moment of the big-rip singularity. 
As a consequence of Eqs. (\ref{beta1}), (\ref{K1}) and (\ref{curv1}) 
the energy density is 
\begin{equation}
\rho = \frac{3 H^2}{8\pi G} \biggl[1 + \frac{\epsilon}{2} a^{-2 - 3\epsilon} 
{\cal B}(\vec{x}) \biggr].
\label{rho1}
\end{equation}
Since $\epsilon > 0$, 
 the relative contribution 
of the gradients to the energy density (second term inside the squared bracket 
of Eq. (\ref{rho1})) and to the extrinsic 
curvature (second term inside the squared bracket in Eq. (\ref{K1}))  
vanishes 
asymptotically,  for $t\to t_{\rm br}$.  For instance, from Eq. (\ref{K1}), 
recalling  Eq. (\ref{HP1}) 
\begin{equation}
 K_{i}^{j}= - H \biggl[ \delta_{i}^{j} - \frac{3\epsilon + 2 }{2}\biggl(\frac{t_{\rm br} - t}{t_{0}}\biggr)^{\frac{2 ( 2 + 3 \epsilon)}{ 3\epsilon}}  {\cal B}_{i}^{j}(\vec{x})\biggr].
 \label{K2}
\end{equation}
 This property of big-rip singularities resembles what is known in the 
 case of usual big-bang singularities where also the gradients 
 decay though with a different power of the cosmic time coordinate.
 Big-bang singularities are argued to be homogeneous 
 but also rather anisotropic \cite{BK1}. As far as the big-rips 
 are concerned, the situation may be very different.
 
 To illustrate the last point, consider a given time $t_{\ast}$ at 
 which gradients can be already neglected. This implies, in the 
 notation of Eq. (\ref{MT}), that 
 \begin{equation}
 \alpha_{xx} = e^{2\,A_{x}(t)},\qquad    \alpha_{yy} = e^{2\,A_{y}(t)},
 \qquad  \alpha_{zz} = e^{2\,A_{z}(t)},
\end{equation}
where, for simplicity, the tensor $\alpha_{ij}$ has been diagonalized 
\footnote{Non-diagonal (or more general) forms of $\alpha_{ij}$ 
do not change the essence of the argument (see below).}.

By solving Einstein equations in the absence of spatial curvature 
it is easy to show that 
\begin{equation}
A_{i}(t) = \frac{2 a(t_{\ast}) {\cal A}_{i}(t_{\ast}) }{ 3 (\epsilon + 2)} 
\biggl[ 1 - \biggl(\frac{a}{a_{\ast}}\biggr)^{-\frac{3}{2}(\epsilon + 2)} \biggr]
\label{ANIS}
\end{equation}
where ${\cal A}_{i}$ are integration constants obeying $\sum_{i} {\cal A}_{i} =0$. Initial conditions have been fixed by requiring that $A_{i}(t_{\ast}) =0$.
Clearly, recalling Eq. (\ref{HP1}), in the limit $t\to t_{\rm br}$ the second 
term in the squared bracket of Eq. (\ref{ANIS}) goes to zero and becomes 
then negligible as the big-rip is approached. In the case of
conventional big-bang singularities, the analog of the second term
in the squared brackets evolves as $a^{3(w -1)/2}$ with $0\leq w < 1$. In this 
case the contribution of the anisotropy clearly grows in the limit $a \to 0$.

 As in the case of conventional big-bang singularities, 
 the evolution of the anisotropy can also be studied in the more  general case when spatial  curvature is included  \cite{BK1}.
 Within our parametrization the contribution of the curvature 
 always decay as $a^{-2}$. Therefore, for $t\to t_{\rm br}$ the 
 r\^ole of the curvature may be neglected. Again, this 
 property is not realized in the vicinity of big-bang singularities. 
 Actually, in the case of big-bang singularities the anisotropy 
 grows as the bang is approached. For more general 
 classes of Bianchi models,  also 
 the spatial curvature grows and this occurrence 
 may induce the celebrated chaotic 
 features and the related BKL oscillations \cite{BK1}. 

Up to now we have considered the case where the big-rip is caused 
by a perfect barotropic fluid with $w < -1$. Needless to say 
that, in the class of models previously investigated, the dominant energy condition is violated. In connection with this problem, both 
in brane-world models \cite{rip5,rip6} and in four-dimensional 
Friedmann-Robertson-Walker models \cite{sudden1,sudden2,sudden3} the possible 
occurrence of a different type of quiescent (or sudden) singularities 
has been emphasized. In sudden (future) singularities the scale 
factor and its first derivative are both finite but higher derivatives of the scale 
factor may diverge\footnote{Along a similar perspective but with 
a different chain of arguments, Ref. \cite{rip7} also argues 
that the presence of a singularity in the future is 
not necessary even if the barotropic index is supernegative.}. In some class of four-dimensional
examples this behaviour implies that while the energy density 
is finite at the rip, the pressure density diverges. 
As correctly pointed out in \cite{sudden1}, sudden singularities 
may occur without a violation of the dominant energy condition.

Consider then the example given in \cite{sudden1} of sudden 
singularities. The scale factor can be written as 
\begin{equation}
a(t) = \biggl(\frac{t}{t_{\rm s}}\biggr)^{q} (a_{\rm s} -1) + 1 - \biggl( 1 -\frac{t}{t_{\rm s}}\biggr)^{n}.
\label{SF2}
\end{equation}
For $ 1 < n < 2$ and $ 0 < q \leq 1$ the solution is defined 
in the interval $0 < t < t_{\rm s}$. For $t \to 0$ the model 
has a curvature (big-bang) singularity where the Hubble 
parameter and the energy density are both divergent. More 
interesting is the second singularity taking place for $t \to t_{\rm s}$. 
For $t\to t_{\rm s}$ the energy density and the Hubble 
parameter are both finite but the pressure density and the second 
time derivative of the scale factor are divergent.

The question we ought to address in the following concerns 
the nature of the gradient expansion in the vicinity 
of sudden singularities.  From Eqs. (\ref{HAM1}) and (\ref{ij1}) 
it can be argued that the contribution of the gradients 
to the quasi-isotropic solution may be different if the scale factor 
behaves as in Eq. (\ref{SF2}): in the limit $t \to t_{\rm s}$ 
both the scale factor and the Hubble rate are finite for the class 
of solutions given in Eq. (\ref{SF2}). On the contrary, as 
previously discussed for big-rips with $w<-1$,
 in the limit $ t\to t_{\rm br}$, the scale 
factor and the Hubble rate are divergent (see Eq.
(\ref{HP1})). This difference is reflected in the contribution 
of the first-order gradients. 

To avoid the proliferation of parameters, the attention will now be focussed on  a particular model belonging to the class defined by Eq. (\ref{SF2}).
 Consider then the case $q= 1/2$ and $n =3/2$. Indeed, the same qualitative results hold for all the models 
 of the class described by Eq. (\ref{SF2}) with the appropriate restrictions mentioned above. Let us also 
 define, for notational convenience the dimensionless variable 
 $\tau = t/t_{\rm s}$. Then a particular solution of the fully 
 inhomogeneous system  including gradients can be 
 easily obtained and expanded for $t\sim t_{\rm s}$, i.e. 
 for $\tau \sim 1$. The most notable difference is that, in this case, pressure 
and energy density are not connected by a barotropic index. 
The final result for the  extrinsic curvature can be written as 
\begin{equation}
K_{i}^{j} = - H(t) \delta_{i}^{j} - \lambda(t) {\cal B}_{i}^{j}(\vec{x}),
\label{exex1}
\end{equation}
whith ${\cal P}_{i}^{j} = t_{\rm s}^{-2} {\cal B}_{i}^{j}$ and  where
 \begin{eqnarray}
 H(t) &=& \frac{1}{2 t_{\rm s}} \frac{ b_{\rm s} + 3\,{\sqrt{\left( 1 - \tau \right) \,\tau}}}
  {\left[ 1 - {\left( 1 - \tau \right) }^{\frac{3}{2}} + b_{\rm s}  \,{\sqrt{\tau}} \right] \,{\sqrt{\tau}}} 
 \nonumber\\
 & \simeq& \frac{1}{4 t_{\rm s} a_{\rm s}^2} [ 
  2 b_{\rm s} a_{\rm s} + 6 a_{\rm s} \sqrt{1 - \tau} + 
  (a_{\rm s} + b_{\rm s} ) b_{\rm s} (1 - \tau) ] + {\cal O}(| 1 - \tau|^{3/2}).
\nonumber\\
\lambda(t) &=&
- \frac{\, 6\,{\sqrt{1 - \tau}} - 3\,\left( -5 + 4\,{\sqrt{1 - \tau}} \right) \,\tau + 10\,
b_{\rm s} \,\tau^{\frac{3}{2}} + 
   6\, \tau^2\,{\sqrt{1 - \tau}}\,  }{15\,t_{\rm s}\,{\left( 1 - {\sqrt{1 - \tau}} + 
   b_{\rm s} \,{\sqrt{\tau}} + 
   {\sqrt{1 - \tau}}\,\tau \right) }^3},
\end{eqnarray}
having defined, for notational convenience, $b_{\rm s} = a_{\rm s} -1$.
As it can be appreciated from the last equations, the exact expressions are rather cumbersome, therefore, in the following, the result for the expansion in the limit $\tau \sim 1$ will be given directly. In particular, 
factorizing $H(t)$  in Eq. (\ref{exex1}) and expanding the relative contribution of the gradients we have       
 \begin{eqnarray}
&&K_{i}^{j} = - H \biggl\{ \delta_{i}^{j} + \biggl[ -\frac{2}{3} \frac{2 a_{\rm s} +1}{b_{\rm s} a_{\rm s}^2} + \frac{2 (2 a_{\rm s} + 1)}{b_{\rm s}^2 a_{\rm s}^2} \sqrt{1 - \tau } 
 \nonumber\\
&& + \frac{(a_{\rm s} + 2) [ 1 + a_{\rm s}^2 ( 4 a_{\rm s} -13) - 10 a_{\rm s}]}{3 b_{\rm s}^{3} a_{\rm s}^3 }  (1 - \tau) + {\cal O}(| 1 - \tau|^{3/2}) \biggr]{\cal B}_{i}^{j}(\vec{x})  \biggr\}.
\label{ec1}
\end{eqnarray}

Clearly, in the limit $ \tau\to 1$ (i.e. $t\to t_{\rm s}$) the contribution of the gradients, weighted by 
the space-dependent factor ${\cal B}(\vec{x})$, is not subleading. This 
behaviour has to be contrasted with the case 
of big rip singularities (see, for instance, Eqs. (\ref{K1}) and (\ref{K2})) 
where for $t \to t_{\rm br}$ the gradient contribution is subleading. 

 In similar the form of $\rho$ and $p$ can also be derived and it is:
 \begin{eqnarray}
&& \rho = \overline{\rho} \biggl\{ 1 - \biggl[\frac{2\,\left( a_{\rm s}^2 - 2 a_{\rm s} - 2\right) \,} {9\,{b_{\rm s}}^2\,a_{\rm s}^2} + \frac{4\,\left( 1 + a_{\rm s} + a_{\rm s}^2 \right) }{3 b_{\rm s}^3\,
 a_{\rm s}^2} \sqrt{1-\tau}
 \nonumber\\
 && + \frac{2\,\left( -2 + 21\, a_{\rm s} + 17\,{{a_s}}^2 + 47\,{{a_s}}^3 - 3\,{{a_s}}^4 + {{a_s}}^5 \right) }
  {9\,{{a_s}}^3\,{{b_s}}^4} (1- \tau)
 + {\cal O}( | 1 -\tau|^{3/2})\biggr] {\cal B}(\vec{x})\biggr\},
 \label{rc1}\\
 && p = \overline{p} \biggl\{ 1 + \biggl[ \frac{\sqrt{1 - \tau}}{3 a_{\rm s}} - 
 \frac{a_{\rm s}^2 -1}{18 a_{\rm s}^2} (1 - \tau) + {\cal O}(|1 - \tau|^{3/2})\biggr] {\cal B}(\vec{x})\biggr\}.
 \label{rp2}
 \end{eqnarray}
 The exact expressions of $\overline{\rho}$ and $\overline{p}$ are a bit 
 involved and then their expansion for $\tau\to 1$ will be given 
 \begin{eqnarray}
 && \overline{\rho} \simeq \frac{3}{32\pi G\,a_{\rm s}^2 \, t_{\rm s}^2}\biggl[ 
 b_{\rm s}^2 + 6 b_{\rm s}  \sqrt{1 - \tau} + \frac{ 2 a_{\rm s}^3 - 5 a_{\rm s}^2 + 13 a_{\rm s} -1}{a_{\rm s}} ( 1 - \tau) + {\cal O}(|1- \tau|^{3/2})\biggr],
 \label{rhobar}\\
 && \overline{p} \simeq \frac{3}{16\pi G a_{\rm s} t_{\rm s}^2 \, \sqrt{1 - \tau} }
 \biggl[ 1 + \frac{a_{\rm s}^2 -1}{2 a_{\rm s}^2}\sqrt{1-\tau} - \frac{3 (a_{\rm s}-1)}{ 2 a_{\rm s}} (1- \tau) + {\cal O}(| 1 - \tau|^{3/2})\biggr].
 \label{pbar}
 \end{eqnarray}
 From Eq. (\ref{rc1}) it can be deduced that as $t \to t_{\rm s}$ the gradients 
 are not subleading. Moreover, from Eq. (\ref{rhobar}) one can also 
 argue that for $t\to t_{\rm s}$, $\overline{\rho}$ is finite. 
 On the contrary, in the same limit, i.e. $\tau \to 1$ the pressure 
 density diverges (see Eq. (\ref{pbar})).  The amusing thing is that, in this case, 
 the relative contribution of the gradients of Eq. (\ref{rp2}) is subleading as 
 $ t\to t_{\rm s}$.  
 
 In conclusion, let us summarize the main findings of the present investigation.
 In the first part it has been shown that if the dominant source of the 
 background geometry is given by a perfect fluid with supernegative barotropic 
 index (i.e. $w < -1$) then the contribution both of the spatial gradients 
 and of the anisotropy tends to decay as the big-rip singularity is approached. 
For the more conventional big-bang singularities the situation is 
a bit different: while gradients also decay in the vicinity of the big-bang, the 
anisotropy and the curvature may well grow and lead to some type 
of chaotic approach to the singularity. 

We then moved to the analysis of sudden singularities. In this 
case the dominant energy condition is not violated. While the scale 
factor, the Hubble parameter and the energy density are all finite as the singularity is approached, the pressure density diverges. In this situation 
we included the contribution of the gradients and showed in an explicit 
example (representative of a more general class of backgrounds) 
that  the relative contribution of the gradients does not decay in the 
vicinity of the sudden singularity.

Various interesting generalizations are left for future works. The analysis 
of single fluid big rip singularities can be generalized to a multi-fluid 
situation in analogy to what recently discussed in the homogeneous 
and isotropic case \cite{fut1}. In this situation it would be also interesting
to discuss the fate of the anisotropy in more general Bianchi models.

\end{document}